%% file: main.tex
\documentclass[superscriptaddress, aps, showpacs, floatfix, prl, twocolumn]{revtex4-2}

\usepackage{graphicx}
\usepackage{dcolumn}
\usepackage{array}
\usepackage{booktabs}
\usepackage{tikz}
\usepackage{amsmath}
\usepackage{bm}
\usepackage{multirow}
\usepackage{bm}
\usepackage{xcolor}

\def\SmFe/{SmFe$_3$(BO$_3$)$_4$}
\newcommand{\wn}{\,cm$^{-1}$\,}
\newcommand{\ff}{\,$4f$-$4f$\,}
\newcommand{\Sm}{Sm$^{3+}$\,}

\begin{document}

\title{Near-infrared magnetic linear and non-reciprocal directional dichroism in the chiral antiferromagnet \SmFe/}

\author{B. Beke}
\email{balint.beke@edu.bme.hu}
\affiliation{Department of Physics, Institute of Physics, Budapest University of Technology and Economics, M\H{u}egyetem rkp. 3., H-1111 Budapest, Hungary}

\author{B. Sz{\'a}sz}
\affiliation{Department of Physics, Institute of Physics, Budapest University of Technology and Economics, M\H{u}egyetem rkp. 3., H-1111 Budapest, Hungary}
\affiliation{Semilab Semiconductor Physincs Laboratory Co. Ltd., Prielle Kornélia u. 2., H-1117 Budapest, Hungary}

\author{I. A. Gudim}
\affiliation{L. V. Kirensky Institute of Physics Siberian Branch of RAS, 660036 Krasnoyarsk, Russia}

\author{D. Szaller}
\affiliation{Department of Physics, Institute of Physics, Budapest University of Technology and Economics, M\H{u}egyetem rkp. 3., H-1111 Budapest, Hungary}
\affiliation{HUN-REN–BME Condensed Matter Physics Research Group, Budapest University of Technology and Economics, M\H{u}egyetem rkp. 3., H-1111 Budapest, Hungary}
\affiliation{Institute of Solid State Physics, TU Wien, 1040 Vienna, Austria}

\author{S.~Bord{\'a}cs}
\email{bordacs.sandor@ttk.bme.hu}
\affiliation{Department of Physics, Institute of Physics, Budapest University of Technology and Economics, M\H{u}egyetem rkp. 3., H-1111 Budapest, Hungary}
\affiliation{HUN-REN–BME Condensed Matter Physics Research Group, Budapest University of Technology and Economics, M\H{u}egyetem rkp. 3., H-1111 Budapest, Hungary}

\begin{abstract}

We investigated the influence of antiferromagnetic (AFM) order on the near-infrared optical response of \SmFe/ by polarization-resolved, magneto-optical spectroscopy. Below the N\'eel temperature, we observed pronounced magnetic linear dichroism at the \ff transitions of Sm$^{3+}$. By rotating the order parameter of the easy-plane AFM ground state using magnetic fields, we demonstrated the magnetic field control of linear dichroism, and deduced polarization selection rules. In addition, we detected non-reciprocal directional dichroism for several \ff transitions. The non-reciprocal absorption appears in the symmetry-allowed toroidal geometry when the in-plane magnetic field is normal to one of the two-fold rotation axes. These results indicate a strong interference between electric- and magnetic-dipole excitations for transitions showing non-reciprocal absorption. The observed optical anisotropies provide a promising path to detect the in-plane orientation of the AFM order, enabling imaging and time-resolved studies of the AFM order parameter in rare-earth ferroborates.

\end{abstract}

\maketitle
\section{Introduction}
Recent advances in spintronics have directed significant interest towards antiferromagnets, which are regarded as a promising platform for robust, high-speed memory devices \cite{wadley_science_2016,jungwirth_natnano_2016,jungwirth_natphys_2018,Krempasky2024}. However, the detection of antiferromagnetic (AFM) order still remains a nontrivial task as they are largely invisible for conventional magnetometric techniques due to the lack of net magnetization. Optical probes can provide a compelling alternative, offering symmetry-sensitive and ultrafast access to magnetic order parameters without the need for large-scale facilities  \cite{nemec_natphys_2018}. 

In non-centrosymmetric antiferromagnets, the magnetoelectric coupling significantly broadens the range of optical phenomena that permit the detection of a compensated AFM order. For example, the electric field-induced magnetization of these compounds can be optically detected via the magneto-optical Faraday (or Kerr-) effect as applied e.g.~in the case of Cr$_2$O$_3$ to visualize AFM domains \cite{Hayashida2022}. A more unusual phenomenon is the non-reciprocal absorption, also termed as non-reciprocal directional dichroism (NDD), which manifests as an absorption difference for opposite light-propagation directions \cite{tokura_jmmm_2007}. NDD has been detected in various transition metal compounds using optical spectroscopy \cite{saito_prl_2008, saito_jpsj_2008}, even close to the telecommunication wavelength \cite{Park2022,Kimura2024,toth_prb_2024}, and also applied for the imaging of AFM domains \cite{saito_nmat_2009,Sato2022,toth_prb_2024}. However, optical magnetoelectric effects have been studied less in rare-earth compounds so far \cite{Smith2026}, despite the mixing of the electric and magnetic dipole excitations giving rise to these phenomena is expected to be effective for \ff transitions \cite{kezsmarki_nc_2014,toyoda_prl_2015}.

\begin{figure}[h]
\centering
\includegraphics[width = 0.7\linewidth]{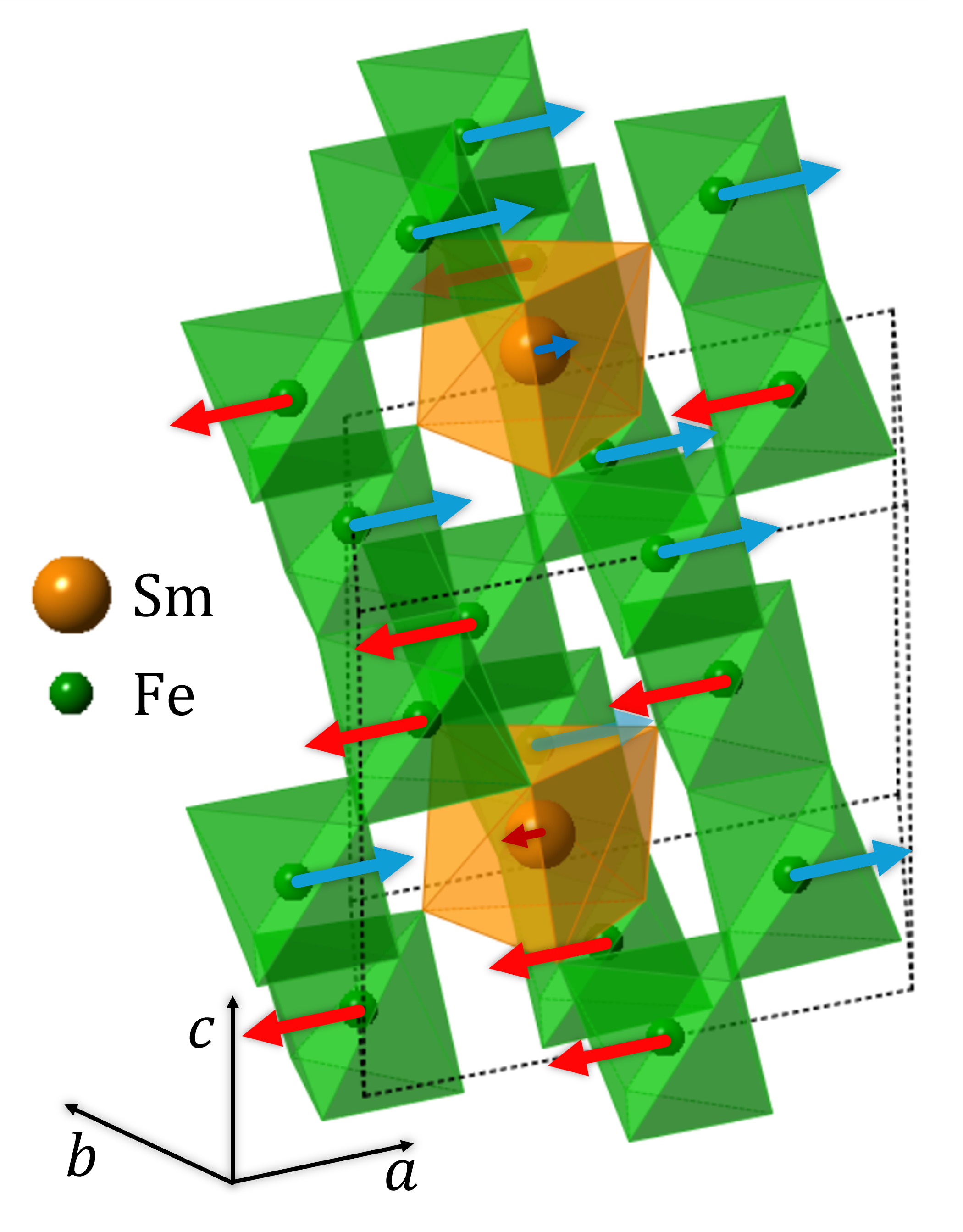}
\caption{Crystal structure and easy-plane antiferromagnetic order in \SmFe/. The mustard and green spheres represent the magnetic Sm$^{3+}$ and Fe$^{3+}$ ions, respectively. The surrounding trigonal-prismatic and octahedral oxygen coordinations are shown in the corresponding colors.  The magnetic moments of the two antiferromagnetically aligned Fe sublattices are indicated by longer blue and red arrows, while the Sm magnetic moments are represented by shorter dark-blue and wine-colored arrows. All moments are shown pointing parallel and antiparallel to the two-fold rotational symmetry axis, $a$. The $c$ axis is the trigonal axis, and the $b$ axis is perpendicular to both $a$ and $c$. The non-magnetic O and B atoms are omitted for clarity. The crystallographic unit cell is indicated by a dashed black rhombohedron.}
\label{fig:fig_structure}
\end{figure}

Here, we investigate magnetic-order-induced optical anisotropies in the collinear antiferromagnet \SmFe/ using infrared magneto-optical spectroscopy. This chiral compound (space group $R32$), belonging to the family of rare-earth iron borates, RFe$_3$(BO$_3$)$_4$ (R = Nd, Sm, Gd, Tb, etc), has attracted considerable attention because its AFM order induces a ferroelectric polarization \cite{ritter_jpcm_2012, popov_jetp_2010,kadomtseva_ltp_2010}. Below T$_N$=34\,K, the S=5/2 spins of Fe$^{3+}$ ions form a collinear AFM structure shown in Fig.~\ref{fig:fig_structure}: within planes perpendicular to the $c$ axis, the spins are parallel, while the sign of the sublattice magnetizations, $\mathbf{m}_1$ and $\mathbf{m}_2$, alternate along $c$. The exchange fields of the Fe sublattices polarize the Sm$^{3+}$ moments. This f-d coupling gives rise to an easy-plane type magnetic anisotropy, which forces the AFM order parameter, $\mathbf{l}=\mathbf{m}_1-\mathbf{m}_2$, to lie within the $ab$ plane. Owing to the absence of inversion symmetry, $\mathbf{l}$ couples to the electric polarization, $\mathbf{P}$ as $(P_a,P_b)\sim(l_a^2-l_b^2,2l_al_b)$ \cite{zvezdin_jetpl_2006}. The net polarization vanishes at zero magnetic field since $\mathbf{l}$ of the AFM domains is randomly oriented within the plane. An in-plane magnetic field $\mathbf{B}$ aligns $\mathbf{l}$ perpendicular to the field, and induces a large electric polarization of 500\,\textmu C/m$^2$ \cite{popov_jetp_2010,kadomtseva_ltp_2010}. Leveraging this polarization, Kuzmenko et al.~demonstrated that also electric fields can control the AFM domains and observed an electric-field-induced change in the intensity of the THz frequency spin excitations \cite{kuzmenko_prl_2018}.

Beyond these static magnetoelectric effects, \SmFe/ also exhibits a pronounced dynamic magnetoelectric response, resulting in strongly nonreciprocal absorption in the THz frequency range \cite{kuzmenko_prb_2015}. 
In \SmFe/, NDD was observed in the toroidal geometry, where an in-plane magnetic field, $\mathbf{B}\parallel b$ induces an electric polarization in the orthogonal direction, $\mathbf{P}\parallel a$ \cite{kuzmenko_prb_2015}. Under these conditions, the absorption coefficient depends on whether light propagates parallel or antiparallel to the toroidal moment, $\mathbf{T}\sim\mathbf{P}\times\mathbf{B}$, highlighting a strong coupling between the magnetic and electric degrees of freedom in the dynamical response. Interestingly, when the magnetic field is applied along the orthogonal direction, $\mathbf{B}\parallel a$, polarized transmission experiments reveal that the spin resonance can be selectively excited either through magnetic-dipole or electric-dipole coupling processes \cite{mukhin_ufn_2015}.

Previous infrared studies on \SmFe/ identified \ff transitions of Sm$^{3+}$ ions \cite{chukalina_pla_2010,popova_jetp_2014}. The free Sm$^{3+}$ ion has a $^6$H$_{5/2}$ ground state and is therefore a Kramers ion. Throughout, we label free-ion multiplets using the spectroscopic notation $^{2S+1}L_J$, where $S$, $L$ and $J$ stand for the spin, orbital and total angular momentum, respectively. The Sm$^{3+}$ ions are coordinated by six oxygen ions forming a twisted triangular prism. This low-symmetry crystal field (site symmetry D$_3$) splits the Sm$^{3+}$ multiplets in \SmFe/. Below the N\'eel temperature, internal magnetic fields associated with the AFM ordering of the Fe sublattices lift the remaining Kramers degeneracy. Comparison of the measured spectra with crystal-field calculations enabled the observed lines to be assigned to specific crystal-field excitations of Sm$^{3+}$ \cite{popova_jetp_2014}.

In this article, we present magnetic-field-dependent infrared spectroscopy of \SmFe/. Our polarization-resolved spectra evidence a significant magnetic-field-induced linear dichroism in \ff transitions of Sm$^{3+}$ ions. The optical anisotropy observed for light polarized in the $ab$ plane can be attributed to the rearrangement of the AFM domains. Furthermore, we observed nonreciprocal absorption in the near-infrared (NIR), close to telecommunication wavelengths. The finite NDD in the toroidal geometry indicates strong electric--magnetic dipole coupling in the $4f$–-$4f$ excitations. 

\section{Symmetry analysis}

\input{magnetoelectric_tensor_table}

Before presenting the experimental results, we analyze how the magnetic symmetries change for a single AFM domain sample as $\mathbf{l}$ rotates in the $ab$ plane. We consider the two principal cases: when $\mathbf{l}\parallel a$ and $\mathbf{l}\parallel b$, i.e.~the AFM vector is parallel and perpendicular to the $C_2$ axis, respectively. 
The results are presented in Tab.~\ref{tab:sym}, which has two main columns corresponding to the zero-field and finite field cases. We label the symmetry operations by the Seitz symbols, $\{R|t\}$, where $R$ denotes a point symmetry or the time-reversal, $\Theta$, and $t$ is a translation. 

In zero field, when the order parameter is oriented along the $a$ axis, $\mathbf{l}\parallel a$, $C_2$  leaves both the magnetic order and the crystal structure intact, whereas the time-reversal operation is a symmetry only in combination with translation $c$, due to the AFM order-induced unit cell doubling along the $c$ axis (see Tab.~\ref{tab:sym}). The presence of the latter, $\{\Theta|c\}$ symmetry, prohibits any linear magnetoelectric effects or NDD in zero magnetic field even for a single-domain sample. However, a magnetic order-induced electric polarization becomes allowed due to the broken inversion symmetry of the chiral crystal structure, and the induced polarization is parallel with the $a$ axis, $\mathbf{P}\parallel a$. 

The application of a magnetic field along $\mathbf{B}\parallel b$ keeps the orientation of the AFM order, $\mathbf{l}\parallel a$, but breaks both of these symmetries and only the combination of the two,  $\{C_2\Theta|c\}$, remains. Thus, several components of the linear magnetoelectric tensor become finite as shown in Tab.~\ref{tab:sym}. Correspondingly, NDD emerges 1) in the toroidal geometry for light propagation along and opposite to the toroidal moment $\mathbf{k}\parallel\mathbf{T}\parallel c$ and 2) in the magnetochiral geometry for propagation along the field, $\mathbf{k}\parallel\mathbf{B}\parallel b$.

When the AFM vector points to the orthogonal direction, $\mathbf{l}\parallel b$, the symmetries $\{C_2|c\}$ and $\{\Theta|c\}$ generate the zero-field magnetic symmetry group. Similarly to the previous case, no linear magnetoelectric effect or NDD is allowed. A magnetic field applied along $\mathbf{B}\parallel a$ keeps the orientation of $\mathbf{l}$, but again several elements of the magnetoelectric susceptibility tensor become allowed in finite fields. However, in this case, NDD may appear only in the magnetochiral geometry, but it is absent for $\mathbf{k}\parallel c$.

\section{Experimental methods}

We measured infrared transmission spectra using a Fourier-transform infrared spectrometer (Varian FTIR 670) and an InGaAs photodetector (Thorlabs, PDA10CS-EC). A $d$=160\,\textmu m thick $ab$-plane cut of \SmFe/ single crystal was glued on the cold finger of an optical cryostat (Janis, ST-300). The relative transmission spectra of the sample were obtained by normalizing the measured spectra to those recorded through a reference aperture. A Janis 9TL-HRTB-30 cryo-free superconducting solenoid provided in-plane magnetic fields, $\mathbf{B}\perp c$, perpendicular to the propagation direction, $\mathbf{k}\parallel c$, realizing the so-called Voigt geometry $\mathbf{k}\perp\mathbf{B}$.

\section{Results}

\begin{figure}[h]
\centering
\includegraphics[width = \linewidth]{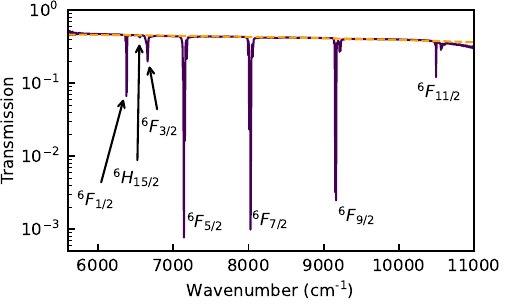}
\caption{Zero-field transmission spectrum of \SmFe/ measured at 5\,K. The orange dashed line marks the transmission baseline due to reflection losses. The symbols labelling the transitions denote the final states of Sm$^{3+}$ \ff excitations.}
\label{fig:fig_transmission}
\end{figure}

A representative transmission spectrum of \SmFe/ measured at 5\,K is shown in Fig.~\ref{fig:fig_transmission}. We observe a series of sharp resonances that we assign to Sm$^{3+}$ \ff excitations in agreement with previous studies \cite{chukalina_pla_2010,popova_jetp_2014}. Following the literature, we label the resonances by the term symbols of the corresponding excited free-ion states of Sm$^{3+}$. Owing to reflections from the sample-vacuum interface, the transmission baseline (orange dashed line) is below one. To account for these reflection losses, we model this baseline using the tail of a high-frequency Lorentzian resonance \cite{Lorentzian}. The relative transmission spectra, \(\mathcal{T}(\omega)\), are normalized against this baseline prior to calculating the absorption coefficient, \(\alpha(\omega) = -\frac{1}{d}\ln \mathcal{T}(\omega)\), where \(\omega \) represents the frequency expressed in wavenumber units (\wn).

\begin{figure*}[t]
    \centering
    \includegraphics[width = \textwidth]{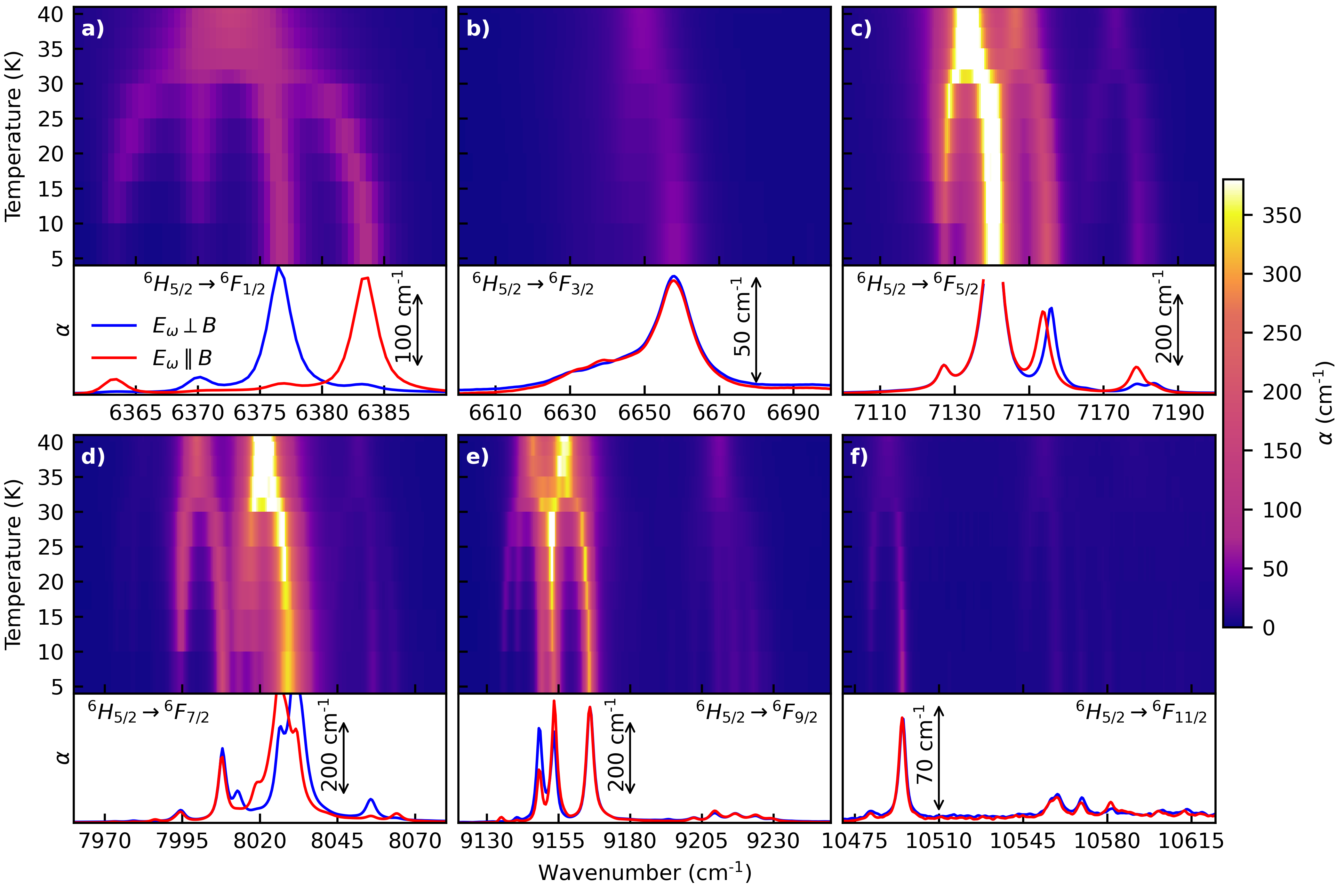} 
    \caption{Temperature dependence of the unpolarized absorption spectrum. Brighter colors in the intensity maps correspond to stronger absorption. The panels below the colormaps show linearly polarized spectra measured in the corresponding frequency ranges at 5\,K in a magnetic field of $B=$2\,T applied along $\mathbf{B}\parallel a$. The incident radiation was polarized either parallel ($\mathbf{E}_\omega\perp\mathbf{B}$, blue curves) or perpendicular ($\mathbf{E}_\omega\parallel\mathbf{B}$, red curves) to the magnetic field.} \label{fig:fig_heatmap_Tdep}
\end{figure*}

To investigate the influence of magnetic ordering on the optical properties of \SmFe/, we measured the temperature dependence of the infrared absorption spectra at low temperatures, across the magnetic ordering. While prior optical spectroscopic studies of this crystal exist, the low-temperature spectra presented there are restricted to the vicinity of the \({}^{6}\)F\({}_{1/2}\)~\cite{chukalina_pla_2010} and \({}^{6}\)F\({}_{11/2}\)~\cite{popova_jetp_2014} transitions. The colormaps in Fig.~\ref{fig:fig_heatmap_Tdep} present the evolution of the unpolarized absorption spectra as a function of temperature.

At temperatures below  T$_N$=34\,K, the spectral lines split due to lifting of the Kramers degeneracy. The low-temperature transition energies are in good agreement with the crystal-field calculations of Popova \textit{et al.}~\cite{popova_jetp_2014}. According to their model, the local exchange field of the antiferromagnetically ordered Fe subsystem splits the ground-state Kramers doublet of \Sm into two levels, $1$ and $1^\prime$, with a zero-temperature energy separation of $\Delta$=13.2\,\wn. The excited $^6$F$_{1/2}$ doublet likewise splits into two levels, denoted $\mathrm{a}$ and $\mathrm{a}^\prime$. Consequently, four transitions are expected around the $^6$F$_{1/2}$ transition at low temperatures as observed in Fig.~\ref{fig:fig_heatmap_Tdep}a. In order of increasing energy, these correspond to $1^\prime\rightarrow \mathrm{a}$, $1^\prime\rightarrow \mathrm{a}^\prime$, $1\rightarrow \mathrm{a}$ and $1\rightarrow \mathrm{a}^\prime$. The two lower-energy branches of the $^6$F$_{1/2}$ excitation, as well as the lower-energy branches of other multiplets, become strongly suppressed below $\sim$10\,K. This behavior can be attributed to the depopulation of the higher-lying initial-state level $1^\prime$, which causes the corresponding optical transitions to vanish.  

Although unpolarized absorption spectra are useful for identifying all optically active resonances, polarized measurements can reveal the corresponding selection rules and thereby provide insight into the microscopic origin of the individual transitions. We investigate the selection rules of \SmFe/ at 5\,K when the AFM domains are aligned by an external magnetic field. The panels below the color maps in Fig.~\ref{fig:fig_heatmap_Tdep} show linearly polarized spectra recorded in a 2\,T magnetic field applied along $\mathbf{B}\parallel a$. The incident radiation was polarized in the $ab$ plane either parallel, $\mathbf{E}_\omega\perp\mathbf{B}$ (blue curves) or perpendicular $\mathbf{E}_\omega\parallel\mathbf{B}$ (red curves) to the field. At such high magnetic fields, the antiferromagnetic order parameter $\mathbf{l}$ is aligned perpendicular to $\mathbf{B}$ within the $ab$ plane. Consequently, the difference between the two absorption spectra corresponds to the magnetic linear dichroism (MLD) of the field-oriented antiferromagnetic domain. The MLD is most clearly observed for the $^6$F$_{1/2}$, $^6$F$_{5/2}$, $^6$F$_{7/2}$ and $^6$F$_{9/2}$ resonances (bottom panels of Figs.~\ref{fig:fig_heatmap_Tdep}a, c, d, e), whereas it is negligible for the $^6$F$_{3/2}$ and $^6$F$_{11/2}$ transitions (Figs.~\ref{fig:fig_heatmap_Tdep}b, f). The MLD-induced polarization contrast is strongest for the $^6$F$_{1/2}$ resonance, enabling a direct determination of the orientation of the antiferromagnetic order parameter.

\begin{figure}[h]
\centering
\includegraphics[width = \linewidth]{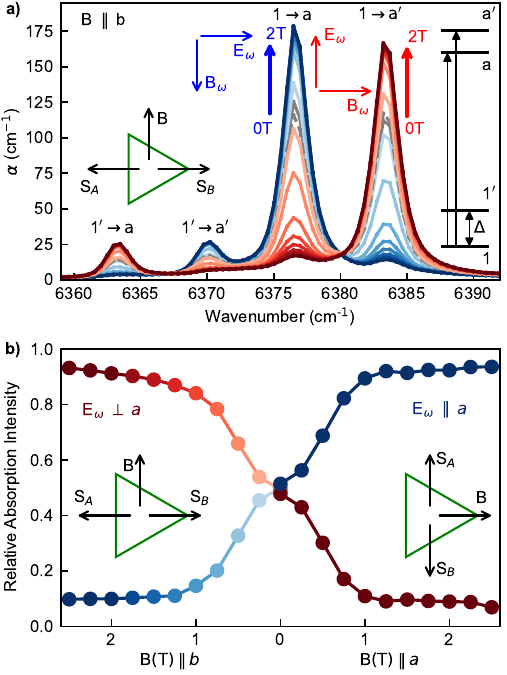}
\caption{Magnetic-field-dependent linearly polarized absorption of the $^6$F$_{1/2}$ resonances at 5\,K. a) Absorption spectra for $\mathbf{E}_\omega\parallel a$ (blue color gradient) and $\mathbf{E}_\omega\parallel b$ (red color gradient). The corresponding zero-field spectra are shown by solid and dashed gray lines, respectively. The magnetic field was aligned along $\mathbf{B}\parallel b$, and the temperature was 5\,K. The level scheme of the $^6$F$_{1/2}$ transition is shown on the right-hand side.  b) Relative absorption intensity of the $1\rightarrow \mathrm{a}^\prime$ (6384 \wn) excitation in the two orthogonal polarizations. The data for $\mathbf{E}_\omega\parallel a$ ($\mathbf{E}_\omega\parallel b$) are displayed on the right-hand side in navy (wine) for a magnetic field along $a$, whereas the corresponding data for $\mathbf{B}\parallel b$ are shown on the left-hand side using color gradients identical to those in a). The orientations of the magnetic field and the magnetic moments of the two sublattices in the trigonal crystal plane are shown in the insets in both panels.} \label{fig:fig_linpol_spectra}
\end{figure}

To explore this optical approach for reading out AFM domains, we investigated the magnetic-field dependence of the linearly polarized absorption at the $^6$F$_{1/2}$ excitation (Fig.~\ref{fig:fig_linpol_spectra}a). The zero-field-cooled spectra measured with light polarization $\mathbf{E}_\omega\parallel a$ (solid gray curve) and $\mathbf{E}_\omega\parallel b$ (dashed gray curve) are nearly identical, and the resonances have an intermediate intensity. When we applied a magnetic field along $\mathbf{B}\parallel b$ the intensity of the transition $1\rightarrow \mathrm{a}$ (6377 \wn) increases (decreases) for light polarization $\mathbf{E}_\omega\parallel a$ ($\mathbf{E}_\omega\parallel b$). A similar phenomenon was observed for the $1\rightarrow \mathrm{a}^\prime$ (6384 \wn) excitation, but the roles of the two polarizations are swapped.

We repeated the experiment with the magnetic field applied along the orthogonal direction within the easy-plane, $\mathbf{B}\parallel a$. Fig.~\ref{fig:fig_linpol_spectra}b compares the magnetic-field dependence of the absorption intensities of the highest-frequency resonance ($1\rightarrow \mathrm{a}^\prime$) from Fig.~\ref{fig:fig_linpol_spectra}a for both magnetic field directions as well as for the orthogonal linear polarizations, $\mathbf{E}_\omega\parallel a$ and $\mathbf{E}_\omega\parallel b$. The data recorded for $\mathbf{B}\parallel a$ are displayed on the right-hand side in navy (wine) for $\mathbf{E}_\omega\parallel a$ ($\mathbf{E}_\omega\parallel b$), whereas the intensities measured for $\mathbf{B}\parallel b$ are shown on the left-hand side using a blue (red) color gradient for $\mathbf{E}_\omega\parallel a$ ($\mathbf{E}_\omega\parallel b$). For both field orientations, the absorption intensities saturate above ~1\,T, with the saturation values depending solely on the relative orientation of the magnetic field and the light polarization. The selection rules observed for the field-aligned  AFM state are complementary for the $1\rightarrow \mathrm{a}$ and $1\rightarrow \mathrm{a}^\prime$ resonances:
$1\rightarrow \mathrm{a}$ is active (silent) for $\mathbf{E}_\omega\perp\mathbf{B}$ ($\mathbf{E}_\omega\parallel\mathbf{B}$), whereas $1\rightarrow \mathrm{a}^\prime$ is active (silent) for $\mathbf{E}_\omega\parallel\mathbf{B}$ ($\mathbf{E}_\omega\perp\mathbf{B}$). Consequently, we conclude that $1\rightarrow a$ is active (silent) for $\mathbf{E}_\omega\parallel\mathbf{l}$ ($\mathbf{E}_\omega\perp\mathbf{l}$), whereas $1\rightarrow \mathrm{a}^\prime$ is active (silent) for $\mathbf{E}_\omega\perp\mathbf{l}$ ($\mathbf{E}_\omega\parallel\mathbf{l}$). 

In addition, we observed the influence of mechanical strain on the linear dichroism. Remounting and regluing the sample in a rotated orientation led to a redistribution of the relative zero-field absorption strengths between the $1\rightarrow \mathrm{a}$ and $1\rightarrow \mathrm{a}^\prime$ resonances without affecting the high-field selection rules \cite{Supplement}. This behavior indicates that the orientation of $\mathbf{l}$ is highly sensitive to the strain induced by the clamping and cooling of the sample. To circumvent this artefact, for the experiments presented in Fig.~\ref{fig:fig_linpol_spectra}b, we rotated the crystal together with the copper sample holder plate to which it had been permanently glued.

\begin{figure}[h]
    \centering
    \includegraphics[width = \linewidth]{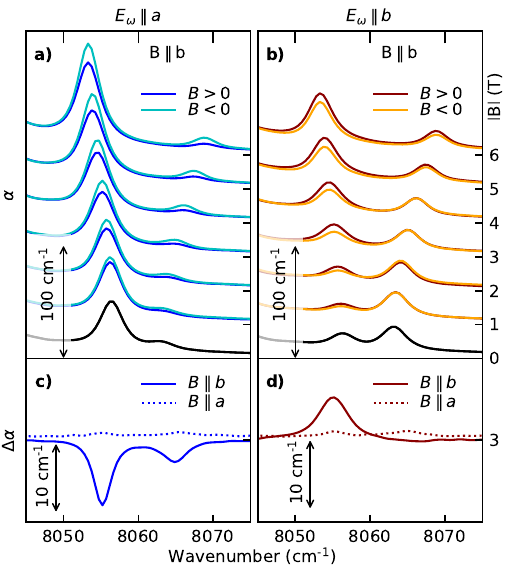}
    \caption{
    Magnetic field dependence of the absorption spectrum at the high energy side of the $^6$F$_{7/2}$ resonances at 5\,K. The two upper panels show absorption spectra with $\mathbf{B}\parallel b$ sample orientation. Darker/lighter colors correspond to positive/negative magnetic field direction. The bottom panels show the difference of the absorption spectra measured in $\pm3$\,T calculated as $\alpha(+\text{B})-\alpha(-\text{B})$. a) and c) panels show the spectra for $\mathbf{E}_\omega\parallel a$, b) and d) panels contain spectra for $\mathbf{E}_\omega\parallel b$.
    } \label{fig:fig_ndd}
\end{figure}

Finally, we report the observation of NDD for \ff transitions in the NIR spectral range. Above $B>$1\,T, where the AFM vector is perpendicular to the field throughout the sample, we detected finite absorption differences for spectra measured in positive and negative magnetic fields when the field was oriented along $\mathbf{B}\parallel b$. This absorption contrast is exemplified for both light polarization directions, $\mathbf{E}_\omega\parallel a$ and $\mathbf{E}_\omega\parallel b$ in Fig.~\ref{fig:fig_ndd}a and b, respectively, for $^6F_{7/2}$ transition. The observed absorption difference is a direct consequence of NDD, since the light propagation is parallel or antiparallel with the magnetic field-induced toroidal moment when the magnetic field direction was reversed. We also found similar NDD signals for transitions $^6F_{5/2}$ and $^6F_{9/2}$ as shown in the Supplement \cite{Supplement}.

Finite NDD can emerge in the infrared range when electric and magnetic dipole excitations can be mixed \cite{kezsmarki_nc_2014,toyoda_prl_2015}. The crystal field calculations of Popova et al.~\cite{popova_jetp_2014} confirmed that the in-plane components of the electric and magnetic dipoles belong to the same irreducible representation, enabling non-reciprocal absorption. The specific strength of the non-reciprocal signal requires a more detailed calculation, which could be an interesting target for future theoretical work.

We further corroborate the direct coupling between the toroidal moment and the NDD by studying the magnetic field orientation dependence of the absorption. We applied a magnetic field also in the orthogonal direction, $\mathbf{B}\parallel a$, where the magnetic-field-induced electric polarization is parallel with the field, and thus, the toroidal moment is zero. As displayed in Fig.~\ref{fig:fig_ndd}c and d, we observed vanishing NDD, implying that the toroidal moment is necessary for finite NDD in agreement with our symmetry analysis in Section II. This result demonstrates that the measurement of the NDD provides another sensitive probe for the in-plane orientation of the AFM vector $\mathbf{l}$. 
Notably, this behavior contrasts with orthophosphates like LiCoPO$_4$ and LiNiPO$_4$, where the sign of $\mathbf{l}$ with respect to the easy axis is detected via NDD in the NIR range \cite{Kimura2024, toth_prb_2024}. 

\section{Discussion}

The magnetic-field evolution of the selection rules presented in Fig.~\ref{fig:fig_linpol_spectra}b resembles that of the sub-THz quasi-AFM magnon of \SmFe/ observed around 9.2\,\wn at 12\,K \cite{kuzmenko_prl_2018}. Since this magnon can be excited by $B_\omega \perp \mathbf{l}$, its absorption intensity tracks the reorientation of $\mathbf{l}$ in the $ab$ plane induced by either magnetic or electric fields. We note that the static electric polarization follows a similar field dependence \cite{popov_jetp_2010, kadomtseva_ltp_2010}.


The common magnetic-field-dependent trends observed from static to dynamic regimes provide a direct insight into the microscopic origin of magnetoelectric effects in \SmFe/. Due to their high atomic number, rare-earth ions exhibit a much stronger spin-orbit coupling than transition-metal ions. Specifically, Fe$^{3+}$ with a half-filled d shell has vanishing orbital moment. Consequently, the field dependence of the electric polarization in \SmFe/ tracks the orientation of the Sm$^{3+}$ moments rather than the Fe$^{3+}$ spins. Because the $g$-factor of Sm$^{3+}$ is small and the samarium magnetic moment is exchange-coupled to the iron sublattice, the Sm$^{3+}$ moments align along $\mathbf{l}$. This mechanism naturally explains why the selection rules of the Sm$^{3+}$ crystal-field transitions, the quasi-AFM magnon, and the macroscopic electric polarization all share a common magnetic-field evolution governed by the orientation of the AFM domains in the $ab$ plane. 
Ultimately, this robust coupling enables both the deterministic electrical control of the magnetic domains \cite{kuzmenko_prl_2018} and their optical readout across disparate length scales, ranging from the millimeter scale in the sub-THz regime \cite {kuzmenko_prl_2018} to the micrometer scale in the NIR range.

\section{Conclusions}

We have investigated the NIR optical response of the chiral antiferromagnet \SmFe/ by polarization-resolved infrared spectroscopy in magnetic fields. We found a pronounced magnetic linear dichroism at the \ff transitions of Sm$^{3+}$ below the Néel temperature. This reflects the breaking of the C$_3$ rotational symmetry of the $ab$ plane due to the AFM ordering. The magnetic-field dependence of the polarization selection rules demonstrates that the optical absorption directly tracks the orientation of the AFM order parameter. Furthermore, we observed NDD for several \ff transitions in the NIR range. The effect is present only in the toroidal geometry and disappears for magnetic-field orientations where the symmetry forbids a finite toroidal moment, in agreement with symmetry considerations. 

Our results extend the family of optical magnetoelectric phenomena in rare-earth ferroborates from the THz to the infrared range \ff transitions, which are close to the telecommunication wavelengths. Beyond revealing the microscopic coupling between AFM order and localized electronic excitations, the observed magnetic linear and non-reciprocal directional dichroisms open a route toward imaging AFM domains and ultrafast optical studies, which may enable optical control of AFM order in magnetoelectric materials.

\section{Acknowledgments}

\begin{acknowledgments}
The authors are grateful to A.~Pimenov and T.~Feh\'er for fruitful discussions, and to S.~Stojanovic for assistance with the sample preparation.
D.~S.~acknowledges the support of the National Research, Development and Innovation Office of Hungary (STARTING 150230, 2023-1.2.4-TÉT-2023-00051).
B.~S. acknowledges the support of the Ministry for Culture and Innovation through the Cooperative Doctoral Programme of the National Research, Development and Innovation Office of Hungary. The Hungarian National Research, Development and Innovation Office, NKFIH also supported this work via grant ADVANCED 152684. The cooperation of Austrian and Hungarian scientists was supported by Austria’s Agency for Education and Internationalisation, OeAD (WTZ HU 02/2024). The authors acknowledge TU Wien Bibliothek for financial support through its Open Access Funding Programme.

I.A.G. provided and characterized the sample. B.B. and S.B. designed and built the optical setup, B.S. adapted the cryostat to the magnet. B.B., B.S. and S.B. performed the experiments. B.B., S.B. and D.S. analyzed the data and B.B. prepared the publication figures. S.B. performed the symmetry analysis. S.B., D.S. and B.B. wrote the manuscript text with contributions from the other authors.   
\end{acknowledgments}

\bibliography{literature}

\end{document}

%% file: magnetoelectric_tensor_table.tex


\renewcommand{\arraystretch}{1.}
\setlength{\tabcolsep}{1pt}

%
%
\newcommand{\tcell}[4][0pt]{%
\begin{minipage}[c][#3][c]{#2}
\centering
\vspace*{#1}%
#4\par
\end{minipage}%
}

%
%
\newcommand{\ltcell}[4][0pt]{%
\begin{minipage}[c][#3][c]{#2}
\raggedright
\vspace*{#1}%
#4\par
\end{minipage}%
}

\newcommand{\dotmatrix}[1]{%
\begin{tikzpicture}[scale=0.38, baseline={(current bounding box.center)}]
  \foreach \i in {1,2,3}{
    \foreach \j in {1,2,3}{
      \fill[gray!30] (\j,-\i) circle (0.08);
    }
  }
  \foreach \p in {#1}{
    \pgfmathtruncatemacro{\r}{int(\p/10)}
    \pgfmathtruncatemacro{\c}{mod(\p,10)}
    \fill[black] (\c,-\r) circle (0.13);
  }
\end{tikzpicture}%
}

\newcommand{\tensormatrix}[1]{%
\begin{tikzpicture}[baseline={(m.center)}]
\node[inner sep=0pt] (m) {$\left[\vcenter{\hbox{\dotmatrix{#1}}}\right]$};
\node[inner sep=0pt, anchor=east] at ([xshift=-0.25em]m.west)
{$\chi^{\mathrm{me}}=$};
\end{tikzpicture}%
}

\newcommand{\HeadH}{2.2em}
\newcommand{\TopH}{2.1em}
\newcommand{\BodyH}{5.2em}
\newcommand{\BlockH}{7.3em}

\begin{table*}[ht]
\centering
\small

\begin{tabular}{|c|c|c|c|c c|}

\hline

&
\multicolumn{2}{c|}{\tcell{4.1cm}{\HeadH}{$\bm{\mathrm{B}}=0$}}
&
\multicolumn{3}{c|}{$\bm{\mathrm{B}} \neq 0$}
\\

\hline


\multirow{2}{*}{%
\tcell{1.8cm}{\BlockH}{%
$\bm{\mathrm{l}} \parallel a$\\
$(\bm{\mathrm{l}}\parallel C_2)$
}}
&
\multirow{2}{*}{%
\tcell{1.7cm}{\BlockH}{%
$\{C_2|0\}$\\
$\{\Theta|c\}$\\
$\{C_2\Theta|c\}$
}}
&
\multirow{2}{*}{%
\tcell{2.2cm}{\BlockH}{%
$\bm{\mathrm{P}} \parallel a$\\
$\chi^{\mathrm{me}} = 0$\\
NDD = $\emptyset$
}}
&
\multicolumn{3}{c|}{%
\tcell{7.8cm}{\TopH}{%
$\bm{\mathrm{B}} \parallel b$
}}
\\

\cline{4-6}

&
&
&
\tcell{1.6cm}{\BodyH}{%
$\{C_2\Theta|c\}$
}
&
\tcell{3.0cm}{\BodyH}{%
$\bm{\mathrm{P}} \parallel a$\\
NDD:
$\bm{\mathrm{k}} \parallel \bm{\mathrm{B}} \parallel b$\\
$\bm{\mathrm{k}} \parallel
\bm{\mathrm{P}} \times
\bm{\mathrm{M}} \parallel c$
}
&
\tcell[-0.4ex]{2.7cm}{\BodyH}{%
\tensormatrix{12,13,21,31}
}

\\

\hline


\multirow{2}{*}{%
\tcell{1.8cm}{\BlockH}{%
$\bm{\mathrm{l}} \parallel b$\\
$(\bm{\mathrm{l}}\perp C_2)$
}}
&
\multirow{2}{*}{%
\tcell{1.7cm}{\BlockH}{%
$\{C_2|c\}$\\
$\{\Theta|c\}$\\
$\{C_2\Theta|0\}$
}}
&
\multirow{2}{*}{%
\tcell{2.2cm}{\BlockH}{%
$\bm{\mathrm{P}} \parallel -a$\\
$\chi^{\mathrm{me}} = 0$\\
NDD = $\emptyset$
}}
&
\multicolumn{3}{c|}{%
\tcell{7.8cm}{\TopH}{%
$\bm{\mathrm{B}} \parallel a$
}}
\\

\cline{4-6}

&
&
&
\tcell{1.6cm}{\BodyH}{%
$\{C_2|c\}$
}
&
\tcell{3.0cm}{\BodyH}{%
$\bm{\mathrm{P}} \parallel -a$\\
NDD:
$\bm{\mathrm{k}} \parallel
\bm{\mathrm{B}} \parallel a$
}
&
\tcell[-0.4ex]{2.7cm}{\BodyH}{%
\tensormatrix{11,22,23,32,33}
}

\\

\hline

\end{tabular}

\caption{Magnetic space group symmetries and their effects on the low-temperature magnetoelectric properties of \SmFe/. Rows correspond to the orientations of the antiferromagnetic vector along the two orthogonal high-symmetry axes of the magnetic easy plane, while columns correspond to zero and finite external magnetic fields. Space group symmetries are given by their Seitz symbols, and $\Theta$ stands for time-reversal. For each configuration, the representative symmetries and the allowed magnetoelectric phenomena, such as the direction of the electric polarization ($P$), the magnetoelectric tensor $\chi^{me}$, and possible experimental geometries of non-reciprocal directional dichroism (NDD), are presented. Symmetry-allowed components of the magnetoelectric tensor are indicated by black dots in the basis of the orthogonal crystallographic axes.} \label{tab:sym}

\end{table*}